\documentclass[aps,prl,twocolumn,showpacs,preprintnumbers,amsmath,amssymb]{revtex4}   

\usepackage{graphicx}
\usepackage{dcolumn}
\usepackage{bm}
\usepackage[dvips]{color}

\newcommand {\beq} {\begin{equation}}
\newcommand {\eeq} {\end{equation}}
\newcommand {\beqa}{\begin{eqnarray}}
\newcommand {\eeqa}{\end{eqnarray}}
\newcommand {\del} {\partial}
\newcommand {\tr}{{\rm tr\,}}

\newcommand {\ee}{\mbox{e}}

\newcommand{\pa}{\partial}

\begin{document}


\title{Higher derivative corrections to black hole thermodynamics\\
from supersymmetric matrix quantum mechanics}
 
\author{Masanori Hanada$^{1}$}
\email{masanori.hanada@weizmann.ac.il}
\author{Yoshifumi Hyakutake$^{2}$}
\email{hyakutake@cep.osaka-u.ac.jp}
\author{Jun Nishimura$^{3,4}$}
\email{jnishi@post.kek.jp}
\author{Shingo Takeuchi$^{4,5}$}
\email{shingo@apctp.org}

\affiliation{
$^{1}$Department of Particle Physics, Weizmann Institute of Science,
Rehovot 76100, Israel\\
$^{2}$Department of Physics, 
Osaka University, Toyonaka, Osaka 560-0043, Japan\\
$^{3}$
High Energy Accelerator Research Organization (KEK), 
Tsukuba, Ibaraki 305-0801, Japan \\
$^{4}$Department of Particle and Nuclear Physics,
School of High Energy Accelerator Science,
Graduate University for Advanced Studies (SOKENDAI),
Tsukuba 305-0801, Japan\\
$^{5}$Asia Pacific Center for Theoretical Physics (APCTP), 
Pohang, Gyeongbuk 790-784, Korea
}

\date{March, 2009; preprint: WIS/19/08-OCT-DPP, OU-HET 612, KEK-TH-1288, APCTP Pre2008-008}


\begin{abstract}
We perform a direct test of
the gauge-gravity duality associated with 
the system of $N$ D0-branes in type IIA superstring theory
at finite temperature.
Based on the fact that higher derivative corrections
to the type IIA supergravity action start at the order
of $\alpha'^{3}$, we derive
the internal energy in expansion around
infinite 't Hooft coupling up to the subleading term
with one unknown coefficient.
The power of the subleading term is
shown to be nicely reproduced by the Monte Carlo data obtained
nonperturbatively on the gauge theory side 
at finite but large effective (dimensionless)
't Hooft coupling constant.
This suggests, in particular, that the open strings
attached to the D0-branes provide the microscopic origin
of the black hole thermodynamics of the dual geometry
including $\alpha '$ corrections.
The coefficient of the subleading term extracted from the fit
to the Monte Carlo data provides a prediction 
for the gravity side, which can be checked
once the complete form of the O$(\alpha'^{3})$ corrections
to the supergravity action 
is obtained.
\end{abstract}



\maketitle


\paragraph*{Introduction.---}

It is widely believed 
that large-$N$ gauge theory provides
a nonperturbative description of 
superstrings \cite{BFSS,IKKT}
and hence of quantum space-time.
In pursuing such a direction,
it is useful to consider
a particular set-up with
a stack of $N$ D-branes in the so-called 
decoupling limit.
After taking this limit there exists a parameter region,
in which the superstring theory in the bulk
ten dimensions reduces to a classical
supergravity theory so that one only has to 
consider a particular classical solution
that describes the $N$ D-branes.
On the other hand,
the worldvolume theory of the $N$ D-branes
is given by a supersymmetric U($N$) gauge theory.
In the above parameter region, which corresponds
to taking the planar large-$N$ limit 
with infinite 't Hooft coupling,
the gauge theory is conjectured
to have a dual description in terms of the
supergravity solution \cite{Maldacena:1997re}.
Including $\alpha '$ corrections on the gravity
side corresponds to including subleading terms
with respect to the inverse 't Hooft coupling constant
on the gauge theory side.
Similarly, including string loop corrections 
corresponds to including $1/N$ corrections.
In fact the gauge theory is well-defined for
arbitrary coupling constant and $N$, and thus
it is expected to be a non-perturbative description of 
superstrings in a certain curved background.

The system of D0-branes in type IIA superstring theory
provides a particularly simple example of 
the gauge-gravity duality \cite{Itzhaki:1998dd}
since the gauge theory in this case lives 
in one dimension, and hence it is nothing but 
matrix quantum mechanics (MQM).
It is also important due to its
connection \cite{BFSS} to M theory \cite{Witten:1995ex}, 
which is expected to emerge in the strong coupling limit of
type IIA superstring theory.
Recently
Monte Carlo studies of the supersymmetric
MQM have been performed \cite{AHNT}
by using a non-lattice 
regularization \cite{Hanada-Nishimura-Takeuchi},
which respects supersymmetry maximally.
The internal energy calculated for a wide range 
of the effective 't Hooft coupling at finite temperature
interpolates nicely 
the weak coupling behavior \cite{HTE},
and the leading asymptotic behavior
in the strong coupling limit predicted 
from the black hole thermodynamics 
of the dual geometry.
Consistent results are obtained
also from the lattice approach \cite{Catterall:2008yz}.
See Refs.\ \cite{KLL} for earlier works on the same system
based on the Gaussian expansion method.



In this Letter 
we perform a precision
test of the above gauge-gravity duality
by considering $\alpha '$ corrections
to the black hole thermodynamics.
%
%
One of our main results is that
%
the internal energy $E$ at temperature $T$ is given as
%
\begin{alignat}{3}
\label{EvsTsub}
  \frac{1}{N^2} \frac{E}{\lambda^\frac{1}{3}} &= c_1
  \Big( \frac{T}{\lambda^\frac{1}{3}} \Big)^\frac{14}{5} 
  - C \Big( \frac{T}{\lambda^\frac{1}{3}} \Big)^\frac{23}{5} \ , \\
\lambda &=  (2\pi)^{-2} \alpha'^{-\frac{3}{2}}g_s N  
\label{gYM2}
\end{alignat}
in the large-$N$ limit with fixed $\lambda \gg T^3$,
where $c_1 \simeq 7.41$.
The first term is known \cite{Itzhaki:1998dd}
from the supergravity analysis \cite{Klebanov:1996un}.
The second term is the one we get
from 
$\alpha '$
corrections,
where $C$ is calculable
once the O$(\alpha'^{3})$ correction
to the supergravity action is obtained completely. 

On the gauge theory side, $\lambda$ corresponds
to the 't Hooft coupling constant.
By comparing Monte Carlo data 
at large but finite $\lambda$ with Eq.\ (\ref{EvsTsub}),
we can test
the gauge-gravity duality including $\alpha '$ corrections.




%
%
%
%


\paragraph*{The dual black hole geometry.---}

The low-energy effective theory of
type IIA superstring theory
can be obtained at the tree level as
$\mathcal{S} = \mathcal{S}_{(0)} + \mathcal{S}_{(1)} + \cdots $
in the $\alpha '$ expansion.
The type IIA supergravity action corresponds 
to the leading term $\mathcal{S}_{(0)}$, which is given
by 
\begin{alignat}{3}
\mathcal{S}_{(0)} 
  &=   
\kappa \!
\int \! d^{10}x  \sqrt{-g} \Big\{ e^{-2\phi} 
\big(R + 4 \partial_\mu \phi \partial^\mu \phi\big) 
- \!  \tfrac{1}{4} G_{\mu\nu} G^{\mu\nu} \Big\} 
\label{eq:action1}
\end{alignat}
%
%
in the string frame, 
where we show the terms which
depend only on the metric $g_{\mu\nu}$, the dilaton $\phi$ and 
the Ramond-Ramond (R-R) 1-form potential $A=A_\mu dx^\mu$
with the field strength $G=dA$.
The coefficient $\kappa$ is given by 
$\kappa^{-1} = 16 \pi  G_{\rm N} = (2\pi)^7 \alpha'^4 g_s^2$
%
in terms of the ten-dimensional Newton constant $G_{\rm N}$.
In the last term of (\ref{eq:action1}), we have
absorbed the tree-level dilaton factor into 
the normalization of the R-R 1-form potential.




In type IIA supergravity,
$N$ D0-branes at finite temperature 
can be described 
by the non-extremal black 0-brane solution.
In the decoupling limit,
we are interested in the excitations of the D0-branes
with fixed energy in the $\alpha ' \rightarrow 0$ 
limit.
Correspondingly, we need to introduce
a new radial coordinate $U=r/\alpha '$
and take the near-horizon limit of
the above solution, which reads \cite{Itzhaki:1998dd}
%
%
\begin{alignat}{3}
  &ds^2 = \alpha' \Big( - \frac{f}{H^\frac{1}{2}} dt^2 
  + \frac{H^\frac{1}{2}}{f} dU^2 
  + H^\frac{1}{2} U^2 d\Omega_8^2 \Big)\ , \label{eq:sugrasol}
  \\
  &e^\phi = \alpha'^{-\frac{3}{2}} H^\frac{3}{4}, 
  \quad G = \alpha'^2 H^{-2} H' dt \wedge dU \ , \notag
\end{alignat}
where $d\Omega_8^2$ represents the line element of $S^8$.
The functions $H(U)$ and $f(U)$ are given as
\begin{alignat}{3}
  &H = \frac{2^4 15 \pi^5 \lambda}{U^7} \ , 
\quad f = 1 - \frac{U_0^7}{U^7} \ ,
\label{eq:Hf1}
\end{alignat}
where $\lambda$
is given by
(\ref{gYM2}).
The metric (\ref{eq:sugrasol}) represents
a black hole geometry with
an event horizon located at $U=U_0$.


\paragraph*{Black hole thermodynamics.---}

Given the black hole geometry, we can 
obtain thermodynamical quantities associated with it
from the geometry at the horizon.

The Hawking temperature $T$ 
is obtained
by requiring that a conical singularity does not
appear at $U=U_0$ 
when one makes the Wick rotation
and compactifies the Euclidean time $\tau=it$ to $\beta=T^{-1}$.
This gives
\begin{alignat}{3}
  T &= \frac{1}{4\pi} H^{-\frac{1}{2}} f' \Big|_{U=U_0} = 
  c_2 \lambda^\frac{1}{3} 
\Big(\frac{U_0}{\lambda^\frac{1}{3}} \Big)^\frac{5}{2} \ , 
\label{eq:T1}
\end{alignat}
where $c_2 = 7/(2^4 15^\frac{1}{2} \pi^\frac{7}{2})$.
Note that the extremal case ($T=0$) corresponds
to choosing $U_0 = 0$.

The Bekenstein-Hawking entropy $S$ is evaluated 
by the area $\mathcal{A}$ of the horizon 
in the Einstein frame as \cite{Klebanov:1996un}
\begin{alignat}{3}
  \frac{S}{N^2} &= \frac{1}{N^2} \frac{\mathcal{A}}{4G_\text{N}} = 
c_3 \Big( \frac{T}{\lambda^\frac{1}{3}} \Big)^\frac{9}{5} \ ,
  \label{eq:S1}
\end{alignat}
where $c_3 = 4^\frac{13}{5} 15^\frac{2}{5} (\pi/7)^\frac{14}{5}$
and we have used Eq.~(\ref{eq:T1}).

The internal energy $E$ is determined by the first law of 
thermodynamics $dE = T dS$,
and it gives the first term in (\ref{EvsTsub}) with 
the coefficient
$c_1 = \frac{9}{14} c_3 = 7.407 \cdots$.
%
%
%
%


Let us recall the region of validity 
for this leading behavior \cite{Itzhaki:1998dd}.
In order for the contributions from string oscillations 
to be neglected, 
the curvature radius $\rho$
of the geometry (\ref{eq:sugrasol})
at $U=U_0$ should be
much larger than the string scale $\sqrt{\alpha'}$,
i.e.,
\begin{alignat}{3}
 \frac{\rho ^2}{\alpha '}
\equiv
 \frac{4\pi^\frac{5}{2} 15^\frac{1}{2} }
{147 
} 
\Big( \frac{\lambda}{U_0^3} \Big)^\frac{1}{2} 
\gg  1  \ ,
\label{eq:curvature}
\end{alignat}
which corresponds to $T/\lambda^{\frac{1}{3}} \ll 1$ due to (\ref{eq:T1}).
%
In order for the string loop effects to be neglected, 
the effective string coupling at $U=U_0$
should be small enough, i.e.,
\begin{alignat}{3}
  g_s e^\phi \equiv 
\frac{2^5 15^\frac{3}{4} \pi^\frac{23}{4}}{N}  
\Big( \frac{\lambda}{U_0^3} \Big)^\frac{7}{4}  \ll 1 \ ,
\label{eq:coupling}
\end{alignat}
which corresponds to $T/\lambda^{\frac{1}{3}} \gg N^{-\frac{10}{21}} $.
%
%
%
%
%

\paragraph*{Higher derivative corrections.---}


%
%
%
%


When (\ref{eq:curvature}) is not met,
we need to consider $\alpha'$ corrections to 
the type IIA supergravity action $\mathcal{S}_{(0)}$.
They can be obtained by calculating tree-level 
scattering amplitudes of the massless modes
in type IIA superstring theory. 

Explicit calculations show that the two-point 
and three-point amplitudes
contribute only to $\mathcal{S}_{(0)}$, and hence 
$\mathcal{S}_{(1)} = \mathcal{S}_{(2)} = 0$.
%
On the other hand, the four-point amplitudes 
are known to give nontrivial contributions to the effective 
action at the order $\alpha'^3$ \cite{GW}.
%

{}From this fact alone, we can deduce the
power of the subleading term in Eq.~(\ref{EvsTsub}).
On dimensional grounds, 
the actual expansion parameter in the $\alpha '$ expansion
is $\alpha ' / \rho^2$, which is the inverse of (\ref{eq:curvature}).
Using (\ref{eq:T1}), this translates to
$(T/\lambda^{\frac{1}{3}})^{\frac{3}{5}}$.
Since the black hole thermodynamics is expected to
receive corrections of the order 
$(\alpha ' / \rho^2)^3 \sim (T/\lambda^{\frac{1}{3}})^{\frac{9}{5}}$,
we obtain
$\frac{14}{5}+\frac{9}{5}=\frac{23}{5}$ as the power of the
subleading term.
%
%
%

\paragraph*{More on $\alpha '$ corrections.---}

Here
we present a more detailed analysis
of the $\alpha '$ corrections,
which yields the power of the
second term in Eq.~(\ref{EvsTsub}).
We hope that our analysis
will be useful
also in calculating the coefficient $C$ 
once the complete form of
$\mathcal{S}_{(3)}$
is obtained.
%

A typical term 
in $\mathcal{S}_{(3)}$
is given by \cite{Ts}
\begin{alignat}{3}
\mathcal{S}_{(3)} 
  & = \kappa
\int d^{10} x \sqrt{-g} 
\Big\{ 
\alpha'^3 e^{-2\phi} 
{\cal R}^4 
+ \cdots \Big\} \ ,
\end{alignat}
%
where ${\cal R}^4$ stands for a scalar quantity obtained
by contracting indices of four Riemann tensors
and multiplying by some numerical factor.
(Its explicit form 
can be found in Ref.~\cite{HO2}, for example.)
Dilaton-dependent terms can be obtained 
by replacing the Riemann tensor by
the second covariant derivative $D^2 \phi$
of the dilaton field \cite{GSl}.
Other possible terms can be written symbolically as 
$\alpha'^3 R^3G^2$, $\alpha'^3 R^2(DG)^2$ and so on \cite{H1,PT}.

The equations of motion 
are derived by taking the variation of 
the effective action 
$\mathcal{S} = \mathcal{S}_{(0)} + \mathcal{S}_{(3)}$
%
%
with respect to $\phi$, for instance, as
\begin{alignat}{3}
  0 &= R + 4 \partial^\mu \phi \partial_\mu \phi 
- 2 e^{2\phi} D_\mu \pa^\mu e^{-2\phi}
+ 
\alpha'^3 
{\cal R}^4 +  \cdot\!\cdot\!\cdot , 
\label{eq:eom2}
\end{alignat}
and similarly for $g_{\mu\nu}$ and $A_\mu$.
Here we assume 
that the solution to these equations
is given by the same form (\ref{eq:sugrasol})
with the functions $H(U)$ and $f(U)$ replaced by
\begin{alignat}{3}
  H = \frac{2^4 15 \pi^5 \lambda}{U^7}(1 +  H_{(3)}) \ , \quad 
  f = 1 - \frac{U_0^7}{U^7} +  f_{(3)}\ .
\label{eq:df}
\end{alignat}
%
Using this Ansatz, Eq.\ (\ref{eq:eom2}) becomes
%
\begin{alignat}{3}
  0 &= - \frac{1}{\alpha'}
  \Big( \frac{U^3}{\lambda} \Big)^\frac{1}{2} \bigg[ \big(1 - \tfrac{U_0^7}{U^7}\big) 
  \big( - 3 U H_{(3)}' + \tfrac{U^2}{2} H_{(3)}'' \big) \label{eq:eom3}
  \\
  &\quad\,
  + \big( 56 f_{(3)} + 16 U f_{(3)}' + U^2 f_{(3)}'' \big) \bigg] 
  + \frac{1}{\alpha'}\Big( \frac{U^3}{\lambda} \Big)^2 h(\tfrac{U_0}{U}) \ . 
  \notag
\end{alignat}
The last term is obtained by substituting
the leading terms of the solution into
the subleading terms in (\ref{eq:eom2}).
The explicit form of $h(\frac{U_0}{U})$ 
can be obtained once $\mathcal{S}_{(3)}$ is given.
It is important that this last term
has an extra factor of $\lambda ^{-\frac{3}{2}}$ compared 
with the other terms, which is understandable
since the effective expansion
parameter is given by $\alpha ' / \rho ^2 \sim
(U_0^3/\lambda)^{\frac{1}{2}}$ as mentioned above.
Note, for instance, that
the fourth term in (\ref{eq:eom2}) is estimated as
$\alpha'^3 \mathcal{R}^4 \sim \alpha'^3
\times (\alpha'^{-1} \lambda^{-\frac{1}{2}})^4
= \alpha'^{-1} \lambda^{-2}$
using $R \sim \alpha'^{-1} \lambda^{-\frac{1}{2}}$
deduced from (\ref{eq:curvature}).
(We have also checked by explicit calculations that
this kind of estimate is true for all possible
subleading terms.)
Since the other equations of motion have the same structure,
we conclude that $H_{(3)}$ and $f_{(3)}$ can be 
written as
\begin{alignat}{3}
   H_{(3)} &\!= 
  \Big(\frac{U_0}{\lambda^\frac{1}{3}} \Big)^{\!\frac{9}{2}} 
  \tilde{H} ( \tfrac{U_0}{U} ) \ , \quad
   f_{(3)} &= \Big(\frac{U_0}{\lambda^\frac{1}{3}} \Big)^{\!\frac{9}{2}} 
  \tilde f ( \tfrac{U_0}{U} ) 
\label{eq:Hf2}
\end{alignat}
with some functions $\tilde H(\frac{U_0}{U})$ and 
$\tilde f(\frac{U_0}{U})$.

%
%
The location of the horizon $U_\text{H}$
is shifted from $U_0$ due to the $\alpha '$ corrections,
and it should be determined from $f(U_\text{H})=0$, which reads
\begin{alignat}{3}
  \frac{U_0}{U_\text{H}} = 1 + \frac{\tilde f(1)}{7} 
\Big(\frac{U_\text{H}}{\lambda^\frac{1}{3}} \Big)^\frac{9}{2} 
\ . \label{eq:horizon}
\end{alignat}
The Hawking temperature is obtained as
\begin{alignat}{3}
  T = \frac{1}{4\pi} H^{-\frac{1}{2}} f' \Big|_{U=U_\text{H}} 
  = c_2 \lambda^\frac{1}{3} 
\Big(\frac{U_\text{H}}{\lambda^\frac{1}{3}} \Big)^\frac{5}{2}
  \Big\{ 1 + c_4 
\Big(\frac{U_\text{H}}{\lambda^\frac{1}{3}} \Big)^\frac{9}{2} \Big\} \ ,
\end{alignat}
where $c_4 = \tilde f(1) - 
\frac{1}{7} \tilde f'(1) - \frac{1}{2} \tilde H(1)$. 
By solving this equation for $U_\text{H}$ iteratively, we obtain
\begin{alignat}{3}
  \frac{U_\text{H}}{\lambda^\frac{1}{3}} &= \Big( 
  \frac{T}{c_2\lambda^\frac{1}{3}} \Big)^\frac{2}{5} 
  \Big\{ 1 - \frac{2}{5} c_4 
\Big( \frac{T}{c_2 \lambda^\frac{1}{3}} \Big)^\frac{9}{5} \Big\} \ .
\end{alignat}

The Bekenstein-Hawking entropy formula
is no longer valid in the presence of higher derivative terms, 
and we need to use the Wald formula~\cite{wald1,wald2}.
For spherically symmetric black holes, it reads
\begin{alignat}{3}
  S &= - 8\pi \int d\Omega_8 
 \frac{\delta \mathcal{S}}{\delta R_{tUtU}} \sqrt{-g_{tt}g_{UU}} 
  \Big|_{U=U_\text{H}} \ , 
\label{eq:wald}
\end{alignat}
where the variation of the action should be taken
by regarding the Riemann tensor as an independent variable.
Explicit calculations yield
\begin{alignat}{3}
  \frac{\delta \mathcal{S}}{\delta R_{tUtU}} &= 
  \frac{\sqrt{-g} e^{-2\phi}g^{tt}g^{UU}}{32 \pi G_\text{N}} 
  \Big\{ 1 + \alpha'^3 (\mathcal{R}^3 + \cdots) \Big\} \ , 
\label{eq:R3}
\end{alignat}
where we define
$\mathcal{R}^3 \equiv 2 g_{tt}g_{UU} 
(\delta \mathcal{R}^4/\delta R_{tUtU})$.
Inserting the leading supergravity solution
to the O($\alpha'^3$) terms in Eq.~(\ref{eq:R3}),
we obtain
$\alpha'^3 (\mathcal{R}^3 + \cdots) = 
s(\frac{U_0}{U})(U/\lambda^\frac{1}{3})^\frac{9}{2}$,
similarly to the argument that led to (\ref{eq:Hf2}).
Therefore
the entropy (\ref{eq:wald}) is evaluated as
\begin{alignat}{3}
  \frac{S}{N^2} &= 
\frac{1}{N^2} \frac{\tilde{\mathcal{A}}}{4G_\text{N}} \Big\{ 1 + 
  s(1) \Big(\frac{U_\text{H}}{\lambda^\frac{1}{3}}\Big)^\frac{9}{2} \Big\} 
  \notag
  \\
  &= c_3 \Big( \frac{T}{\lambda^\frac{1}{3}} \Big)^\frac{9}{5}
  \Big\{ 1 + c_5 
\Big( \frac{T}{c_2 \lambda^\frac{1}{3}} \Big)^\frac{9}{5} \Big\} \ ,
\end{alignat}
where $c_5 = - \frac{9}{5}c_4 + s(1)$
and the horizon area $\tilde{\mathcal{A}}$
includes $\alpha '$ corrections
through Eqs.~(\ref{eq:Hf2}) and (\ref{eq:horizon}).
The internal energy is obtained as
(\ref{EvsTsub}), where $C = - \frac{28}{23} c_1 c_5 (c_2)^{-\frac{9}{5}} $.
%



\paragraph*{The worldvolume theory.---}

The worldvolume theory of $N$ D0-branes
is given by the U($N$) supersymmetric 
MQM defined by the action
\beqa
S
&=& 
\frac{N}{\lambda} \int
_0^{\beta}  
d t \, 
\tr 
\bigg\{ 
\frac{1}{2} (D_t X_i)^2 - 
\frac{1}{4} [X_i , X_j]^2  
\nonumber \\
&~& 
+ \frac{1}{2} \psi_\alpha D_t \psi_\alpha
- \frac{1}{2} \psi_\alpha (\gamma_i)_{\alpha\beta} 
 [X_i , \psi_\beta ]
\bigg\} \ ,
\label{cQM}
\eeqa
where $D_t  = \del_t
  - i \, [A(t), \ \cdot \ ]$ represents the covariant derivative
with the gauge field $A(t)$ being an $N\times N$ Hermitian matrix.
The model describes
the open string degrees of freedom attached to the branes,
which are decoupled from the bulk degrees of freedom
in the decoupling limit $\alpha ' \rightarrow 0$
with fixed $\lambda$. Note that $N$ and $\lambda$ can be
arbitrary for this statement.
\paragraph*{Monte Carlo results.---}

In simulating the model (\ref{cQM}),
we fix the gauge by the static diagonal gauge
$A(t) = \frac{1}{\beta} {\rm diag}
(\alpha_1 , \cdots , \alpha_N)$,
where $- \pi < \alpha_a \le \pi$,
%
%
%
and introduce a UV cutoff $\Lambda$ as
$X_i ^{ab} (t) = \sum_{n=-\Lambda}^{\Lambda} 
\tilde{X}_{i n}^{ab} \ee^{2 \pi i n t/ \beta}$.
%
%
%
%
%
%
Integration over the fermionic matrices
yields a complex Pfaffian,
which is replaced by its absolute value
following
the argument in Ref.\ \cite{Hanada:2008gy} based on
the large-$N$ factorization.

\begin{figure}[htb]
\begin{center}
\includegraphics[height=6cm]{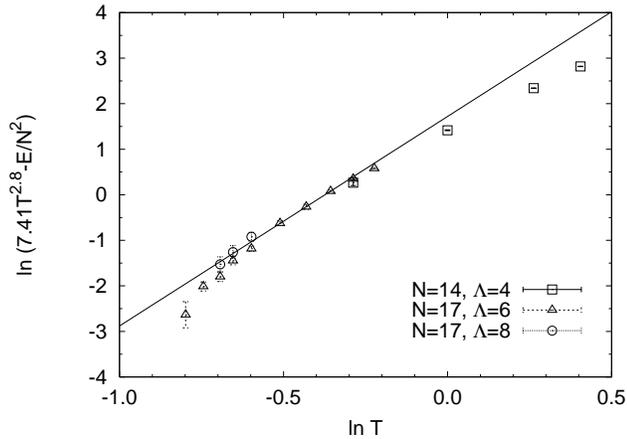}
\end{center}
\caption{
The deviation of the internal energy
$\frac{1}{N^2} E$ from the leading term
$7.41 \, T^{\frac{14}{5}}$
is plotted against the temperature
in the log-log scale for $\lambda=1$.
The solid line represents 
a fit to a straight line with the slope 4.6
predicted from the $\alpha '$ corrections 
on the gravity side.
}
\label{energy}
\end{figure}

\begin{figure}[htb]
\begin{center}
\includegraphics[height=6cm]{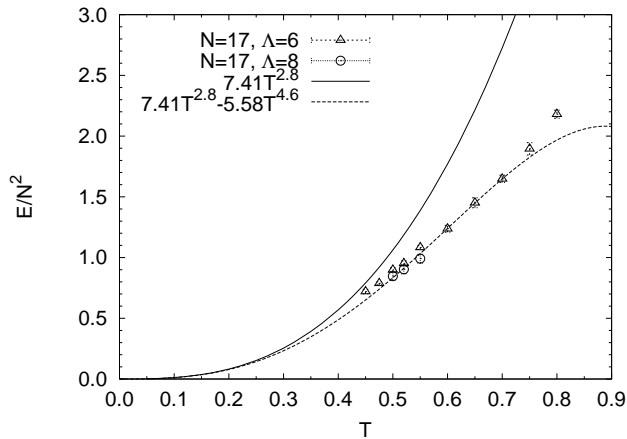}
\end{center}
\caption{
The internal energy $\frac{1}{N^2} E$
is plotted against $T$ for $\lambda=1$.
The solid line represents
the leading asymptotic behavior at small $T$
predicted by the gauge-gravity duality.
The dashed line represents
a fit to the behavior (\ref{EvsTsub})
including the subleading term
with $C=5.58$.
}
\label{energy2}
\end{figure}

The effective coupling constant
is given by $\lambda_{\rm eff}\equiv \lambda/T^3$,
and we set $\lambda=1$ in actual simulations
without loss of generality.
%
In Fig.\ \ref{energy} we plot 
$7.41 \, T^{\frac{14}{5}} - \frac{1}{N^2} E $
against
$T$ in the log-log scale.
Indeed the plot reveals a clear power-law
behavior of the sub-leading term with the power
$\frac{23}{5}=4.6$ as predicted in Eq.\ (\ref{EvsTsub}).
%
The data points for $\Lambda=6$ show
small discrepancies at $T \lesssim 0.55$, 
which can be understood as finite $\Lambda$ effects 
by comparing them with the data points for $\Lambda=8$.
Fig.\ \ref{energy2} shows 
a linear plot of the energy as a function of $T$.
Fitting the data within $0.5 \le T \le 0.7$ 
(with largest $\Lambda$ at each $T$)
to $\frac{1}{N^2} E = 7.41 \, T^{\frac{14}{5}} - C \, T^{p}$,
we obtain $p=4.58(3)$ and $C=5.55(7)$.
If we instead make a one-parameter fit with $p=4.6$ fixed,
we obtain $C=5.58(1)$.
This value, in turn, provides a prediction for the $\alpha'$ corrections
on the gravity side.
%
%

%

%
%
%
%



\paragraph*{Summary.---}

We have discussed the $\alpha '$ corrections to the
black hole thermodynamics,
which enable us to 
determine the power of 
the sub-leading term in (\ref{EvsTsub}).
This power is then found to be reproduced precisely
by Monte Carlo data in gauge theory.
Let us emphasize that the subleading term
is crucial for the precision test of the gauge-gravity duality.
It is intriguing that 
our results in gauge theory 
can tell us
the absence of O($\alpha '$) and O($\alpha '^{2}$) corrections
to the supergravity action.

Recently \cite{Hanada:2008gy}
Monte Carlo data for the Wilson loop 
are also shown to reproduce a prediction obtained by estimating
the disk amplitude in the dual geometry.
Unlike the present case, 
$\alpha '$ corrections to that quantity
start at O($\alpha '$) 
due to the fluctuation of the string worldsheet and its coupling
to the background dilaton field.

While it is certainly motivated to obtain the coefficient $C$
of the subleading term
from gravity,
our results
already 
provide a strong evidence 
that the
gauge-gravity duality holds including $\alpha '$ corrections.
This, in particular, implies that we can understand
the microscopic origin of the black hole thermodynamics
\emph{including $\alpha '$ corrections}
in terms of the open strings attached to the D0-branes.


\paragraph*{Acknowledgments.---}
The authors would like to thank 
O.~Aharony, K.N.~Anagnostopoulos and 
A.~Miwa for discussions.
The computations were carried out on
supercomputers (SR11000 at KEK)
as well as on PC clusters at KEK and Yukawa Institute. 
The work of J.N.\ and Y.H.\ is supported in part by Grant-in-Aid
for Scientific Research (Nos.\ 19340066, 20540286 and 19740141).
%
%



\end{document}